\documentclass[twocolumn]{aastex631}
\usepackage{amsmath}

\begin{document}

\title{A reference meteor magnitude for intercomparable fluxes}
\author{Althea V.\ Moorhead}
\email{althea.moorhead@nasa.gov}
\affiliation{NASA Meteoroid Environment Office, Marshall Space Flight Center, Huntsville, Alabama 35812, USA}

\author{Denis Vida}
\affiliation{Department of Physics and Astronomy, University of Western Ontario, London, Ontario N6A 3K7, Canada} 
\affiliation{Institute for Earth and Space Exploration, University of Western Ontario, London, Ontario N6A 5B8, Canada}

\author{Peter G.\ Brown}
\affiliation{Department of Physics and Astronomy, University of Western Ontario, London, Ontario N6A 3K7, Canada} 

\author{Margaret D.\ Campbell-Brown}
\affiliation{Department of Physics and Astronomy, University of Western Ontario, London, Ontario N6A 3K7, Canada}

\begin{abstract}

The rate at which meteors pass through Earth's atmosphere has been measured or estimated many times over; existing flux measurements span at least 12 astronomical magnitudes, or roughly five decades in mass. Unfortunately, the common practice of scaling flux to a universal reference magnitude of +6.5 tends to collapse the magnitude or mass dimension. Furthermore, results from different observation networks can appear discrepant due solely to the use of different assumed population indices, and readers cannot resolve this discrepancy without access to magnitude data. We present an alternate choice of reference magnitude that is representative of the observed meteors and minimizes the dependence of flux on population index. We apply this choice to measurements of recent Orionid meteor shower fluxes to illustrate its usefulness for synthesizing independent flux measurements.

\end{abstract}

\section{Introduction}
\label{sec:intro}

Historically, meteor shower activity has usually been measured in terms of zenithal hourly rate (ZHR), the number of meteors observed per hour by a visual observer under ideal conditions \citep{norton08}. The ZHR is related to but distinct from meteor flux \citep{rendtel2022}. In recent decades, however, a number of research groups have begun to report meteor fluxes from camera networks or meteor radars \citep[some examples include][]{cb04,blaauw16,molau22,vida22}. These networks vary in sensitivity and coverage, and thus measure the meteor flux in different size regimes. For instance, the Southern Argentina Agile MEteor Radar (SAAMER) measures meteor fluxes down to a limiting equivalent optical magnitude of +9 \citep{bruzzone21}, while the NASA All-Sky Fireball Network has been used to measure the Perseid meteor shower flux to a limiting magnitude of approximately -3.5 \citep{ehlert20}.

If the flux of a single meteor shower is measured simultaneously by several different networks with different limiting magnitudes, these fluxes can be combined to inspect the shower's mass (or magnitude) distribution. For instance, \cite{cb16} combined flux data from five different optical systems as well as the Canadian Meteor Orbit Radar (CMOR) in order to construct a mass distribution that spanned four decades in mass. Similarly, \cite{blaauw17} combined radar, optical, and lunar impact data to construct a mass profile for the Geminids. 

The determination of a network's limiting magnitude, however, is not straightforward, as the detection limit usually varies over the detector's field of view (FOV). For example, the distance from which meteors are viewed tends to vary within the FOV in a way that depends on FOV width and orientation; if the region surveyed by a portion of the FOV is more distant, the meteor must be brighter to be detected. One could address this by discarding all meteors that are too dim to be detected in any part of the field of view, but this typically results in a large reduction in the overall count. 

\begin{figure}
    \centering
    \includegraphics{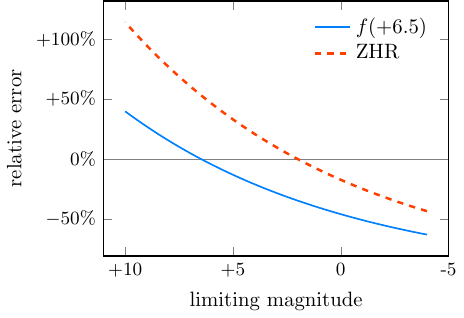}
    \caption{Relative error in magnitude-6.5-limited flux and ZHR computed using an assumed population index of ${r=2.0}$, assuming that the true population index is ${r=2.2}$. This plot covers limiting magnitudes ranging from those of patrol radars to fireballs. }
    \label{fig:rerr}
\end{figure}

The most common way to deal with a variable limiting magnitude is to divide the FOV into smaller pieces and scale the area of each piece to reflect its contribution to the overall flux, assuming that meteor magnitudes follow a power-law distribution \citep{Kaiser1960,koschack90a,bellot94b,vida22,molau22}. This is known as the effective collection area method and is described in Section~\ref{sec:eca}. Because the areas are scaled to correspond to a constant reference limiting magnitude, one can in theory choose any reference magnitude one likes, so long as the power-law assumption holds over the range of magnitudes considered. A common choice is to scale the flux to a constant limiting magnitude of +6.5 and/or convert to ZHR \citep[e.g.,][]{roggemans89,bruzzone21,molau22,vida22} to facilitate comparisons with visual rates \citep{koschack90b}:
\begin{align}
    f(+6.5) &= f(M_\text{ref}) \, 
        r^{6.5 - M_\text{ref}} 
        \label{eq:fm} \\
    \text{ZHR} &= \frac{f(+6.5) \cdot 37\,200~\text{km}^2}{(13.1 \, r - 16.5)(r-1.3)^{0.748}} \label{eq:zhr}
\end{align}
We use $f(M)$ to denote the flux of meteors with absolute magnitudes brighter than $M$, and we use $M_\text{ref}$ to refer to the chosen reference magnitude.
Notice that the above conversion depends on the assumed population index, $r$; if the flux is based on magnitudes that are very different from +6.5, the result can be quite sensitive to the choice of $r$ (see Figure~\ref{fig:rerr}). For example, if the meteor flux measured by CMOR is scaled from its limiting magnitude of +8 to an equivalent ZHR, but ${r=2.0}$ is used for a shower whose true value is ${r=2.2}$, the ZHR will be inflated by 77\%.
Note that these fluxes include all meteors brighter than the given limiting magnitude. This is distinct from the dimmest apparent magnitude visible to a visual observer in the center of their visual field, which is also sometimes called the limiting magnitude \citep[e.g.,][]{koschack90b}, but which we call the visual limit.

Thus, while the use of a common limiting magnitude allows us to compare fluxes published by different groups using different data, it obscures the characteristic limiting magnitude of the system ($M_\text{ref}$) and is sensitive to errors in the assumed magnitude distribution. If two independent flux measurements disagree, it can be unclear whether this is because the data are truly in conflict or because an incorrect choice of $r$ has been used. Even when it is clear that two sets of measurements use different values of $r$, it is usually impossible for the reader to invert that choice unless magnitude information is also provided (or, in the case of the Video Meteor Network, a tool is provided that allows the reader to specify $r$; see Section~\ref{sec:vmn}). For this reason, the \cite{cb16} and \cite{blaauw17} studies exclusively used data from their own observation networks, which allowed them to enforce a consistent set of assumptions.

One possible solution to this problem of incompatible flux measurements is the choice of a reference magnitude that minimizes the dependence of the measured flux on population index. We derive an expression for such a reference magnitude in Section~\ref{sec:mref} that is compatible with the widely used ``effective collection area'' method of measuring fluxes. In Sections~\ref{sec:mavg} and \ref{sec:mavg2}, we provide two alternative ways to estimate $M_\text{ref}$; these estimates can be used by researchers to validate their flux calculations, and may in some cases enable readers to estimate $M_\text{ref}$ for published fluxes. In Section~\ref{sec:ori}, we apply these equations to several independent measurements of Orionid fluxes to demonstrate their value for synthesizing disparate measurements.

\section{Reference limiting magnitude}

The most common approach used to derive meteor flux from observed counts is the effective collection area (ECA) method. This method appears to have originated with \cite{koschack90a} and \cite{koschack90b}, who used a similar approach to estimate the number of meteors that can be detected by a visual observer. \cite{bellot94b} adapted the approach to compute fluxes from camera data. These efforts were predated by \cite{Kaiser1960}, who developed an equivalent approach for radar data.

Today, it is widely used within the meteor astronomy community \citep[recent examples include][]{blaauw17,margonis19,molau22,vida22}.
We review this method in Section~\ref{sec:eca}. We then derive a reference magnitude in Section~\ref{sec:mref} that minimizes the dependence of flux on population index.

We do not specify a method for determining the limiting magnitudes of a meteor observation network; this will depend on the properties of the network and the meteor detection algorithm. Our assumptions are that the limiting magnitude of the network has been characterized and that the ECA method is being used to measure fluxes. We refer readers to \cite{molau16} and \cite{vida22} for more information about the calculation of optical limiting magnitudes, and to \cite{Kaiser1960} for an analogous discussion of radar fluxes. We do, however, offer an alternate method for estimating the reference magnitude that can be used to test whether the limiting magnitudes have been correctly calculated (see Section~\ref{sec:mavg2}).

\subsection{The effective collection area (ECA) method}
\label{sec:eca}

Let $f(M_\text{ref})$ denote the flux of meteors brighter than reference magnitude $M_\text{ref}$. The observed number of meteors follows a Poisson distribution with an expected value of
\begin{align}
    \langle N(M_\text{ref}) \rangle &= f(M_\text{ref}) \, (\vec{A} \cdot \hat{v}) \, T  \, ,
\end{align}
where $N(M_\text{ref})$ is the number of meteors brighter than magnitude $M_\text{ref}$ passing through area $\vec{A}$ within time interval $T$. If the meteors belong to a shower, they share a common velocity vector $\vec{v}$; the unit vector $\hat{v}$ is aligned with the meteors' velocity vector. Note that $\vec{v}$ is the meteor's motion through space, and not its apparent angular motion; the latter plays a role in determining the limiting magnitude \citep{kingery17,molau16}.

If the collection area is flat and its normal vector points toward zenith, then the area presented to the shower is
\begin{align}
    \vec{A} \cdot \hat{v} &= \cos Z_R \, ,
\end{align}
where $Z_R$ is the angle between the local zenith and the shower radiant (also known as the zenith angle). However, meteors entering the atmosphere at a high angle of incidence encounter a gentler atmospheric density gradient along their path, which is thought to decrease their peak brightness \citep{zvolankova83}. Both \cite{zvolankova83} and \cite{molau13} suggest that this can be handled by applying an exponent to the zenith angle term:
\begin{align}
\cos^\gamma Z_R \, ,
\end{align}
where the constant $\gamma$ varies between showers but is approximately 1.5-2.

As discussed in Section~\ref{sec:intro}, the limiting meteor magnitude typically varies within the field of observation. The ECA method handles this by dividing the field of view into a grid and assuming that the limiting magnitude does not vary appreciably within a grid cell. If cell~$i$ covers area $A_i$ and has limiting meteor magnitude $M_i$, then the expected number of meteors in that cell is
\begin{align}
    \langle n_i \rangle &= f(M_i) \, A_i \, T  \cos^\gamma\!Z_R \, .
\end{align}

Recall from equation~\ref{eq:fm} that the relationship between limiting magnitude and flux is a power law; that is,
\begin{align}
    f(M_i) &= f(M_\text{ref}) \,  r^{M_i - M_\text{ref}} \, ,
    \label{eq:fi}
\end{align}
where $r$ is the population index. Therefore,
\begin{align}
    \langle n_i \rangle &= f(M_\text{ref}) ~ r^{M_i - M_\text{ref}} ~ A_i \, T \cos^\gamma\!Z_R \, .
    \label{eq:ni}
\end{align}

The sum of a set of independent Poisson random variables is itself a Poisson random variable, and its expected value is the sum of the expected values of its components. Thus, the expected total number of observed meteors across all cells is
\begin{align}
    \langle N \rangle &= \sum_i \langle  n_i \rangle \nonumber \\
    &= f(M_\text{ref}) \, T  \, \left( \sum_i A_i \, r^{M_i - M_\text{ref}} \, \cos^\gamma\!Z_R \right) \, ,
    \label{eq:nobs}
\end{align}
where the quantity in parentheses is called the effective collection area:
\begin{align}
    A_\text{eff} &= \sum_i A_i \, r^{M_i - M_\text{ref}} \, \cos^\gamma\!Z_R \, .
    \label{eq:aeff}
\end{align}
Equation~\ref{eq:nobs} can then be rewritten as
\begin{align}
    \langle N \rangle &= 
    f(M_\text{ref}) \, A_\text{eff} \, T  \, .
\end{align}

The observed number of events, ${N = \sum_i n_i}$, is an unbiased estimate for the rate of the Poisson process. Therefore, we estimate the flux as
\begin{align}
    f(M_\text{ref}) &\approx \frac{N}{A_\text{eff} \, T}
    \label{eq:fest}
\end{align}
and recover (for example) Equation~1 of \cite{molau22}.

Note that if $M_i$ varies smoothly across the FOV, one could replace the sum in Equation~\ref{eq:aeff} with an integral. For instance, \cite{Kaiser1960} integrates the equivalent of $r^{M_i}$ along the radar echo line; see Equation~18 of that paper. However, most optical applications of the ECA methods use the discrete form presented here.

\subsection{Derivation of reference magnitude}
\label{sec:mref}

We seek a reference limiting magnitude that minimizes the dependence of flux on population index. To obtain this, we set
\begin{align}
    \frac{d f}{d r} &= - \frac{N}{T}
    \frac{1}{A_\text{eff}^2} \frac{d A_\text{eff}}{d r} = 0
\end{align}
It can be shown that ${d A_\text{eff}/d r}$ is zero when:
\begin{align}
    M_\text{ref} &= \frac{\sum_i M_i \, A_i \, r^{M_i} }{\sum_i A_i \, r^{M_i}}
    \label{eq:mref}
\end{align}
We see from this equation that the optimal reference magnitude is a weighted average of the limiting magnitudes of each cell, and therefore lies within the network's limiting magnitude range.

If we multiply both the numerator and denominator of equation~\ref{eq:mref} by $f(M_\text{ref})$, ${r^{-M_\text{ref}}}$, ${\cos^\gamma\!Z_R}$, and $T$, we find that can re-write equation~\ref{eq:mref} as follows:
\begin{align}
    M_\text{ref} 
&= \frac{\sum_i M_i \, \langle  n_i \rangle}{\sum_i \, \langle n_i \rangle} 
\label{eq:mexpn}
\end{align}
We see that the weights are proportional to the expected number of meteors in each cell. Thus, $M_\text{ref}$ is a weighted average of the limiting magnitude \citep[which is used by][]{vida22}, where the weights are the expected rates for each cell.

\subsection{Average limiting magnitude per meteor}
\label{sec:mavg}

The form of equation~\ref{eq:mexpn} hints that we can estimate the reference magnitude using the observed meteor counts:
\begin{align}
    M_\text{ref} &\approx \frac{\sum_i M_i \, n_i}{\sum_i n_i} 
    \label{eq:mn}
\end{align}
Notice that we have replaced the expected counts, $\langle n_i \rangle$, with the observed counts, $n_i$. This approximation makes it clear that our proposed reference magnitude not only lies within the network's range of limiting magnitudes, but is in fact the per-meteor mean limiting magnitude. This makes $M_\text{ref}$ a doubly appealing choice in that it is representative of the limiting magnitude distribution in addition to minimizing the dependence on $r$.

We suggest that researchers publishing fluxes verify that the approximation given in equation~\ref{eq:mn} is similar to the true value given by equation~\ref{eq:mref}. A large discrepancy could indicate that either the cell areas or population index are incorrect. We will also show in an upcoming paper that equation~\ref{eq:mn} provides a reference magnitude that minimizes the covariance between flux and population index when fitting for both quantities simultaneously.

If we use the sample mean of the limiting magnitude (Eq.~\ref{eq:mn}) as a reference magnitude, then a sensible extension is to use the sample standard deviation of the limiting magnitude to describe the range of limiting magnitudes within a meteor detection network:
\begin{align}
    s_\text{ref} &= \frac{\sum_i n_i (M_i - M_\text{ref})^2 }{\left( \sum_i n_i \right) - 1}
\end{align}
If the distribution of limiting magnitudes is strongly skewed or multi-model, researchers could consider quoting another measure of variation, such as the interquartile range.

\subsection{Average absolute magnitude}
\label{sec:mavg2}

If we assume that $M_i$ places a strict upper limit on the absolute meteor magnitude that can be observed in cell $i$, then the probability of observing a meteor with absolute magnitude $M'$ in that cell is:
\begin{align}
    P(M' \, | \, M_i) &= \ln r \times \begin{cases}
        r^{M' - M_i} & M' < M_i \\
        0 & M' > M_i
    \end{cases}
\end{align}
Using this probability distribution, we can calculate the average absolute magnitude we expect for meteors in that cell:
\begin{align}
    \langle M' \, | \, M_i \rangle &= \ln r \int_{-\infty}^{M_i}
    M' \, r^{m - M_i} \, d M'
    = M_i - \frac{1}{\ln r}
    \label{eq:expmm}
\end{align}
We see that the average observed magnitude in any given cell should be systematically offset from the limiting magnitude in that cell by a constant value of $-1/\ln r$. 
Furthermore, the probability that a meteor is detected in a cell with limiting magnitude $M_i$ is:
\begin{align}
    P(M_i) &= \frac{\langle n_i \rangle}{\langle N \rangle} = \frac{A_i \, r^{M_i}}{\sum_j A_j \, r^{M_j}}
    \label{eq:pmi}
\end{align}
We can now combine equations~\ref{eq:expmm} and \ref{eq:pmi} to obtain the expected absolute magnitude of a meteor detected by the system:
\begin{align}
    \langle M' \rangle &= \sum_i \langle M' \, | \, M_i \rangle \, P(M_i) 
    = M_\text{ref} - \frac{1}{\ln r}
    \label{eq:expm}
\end{align}
If we substitute the mean absolute meteor magnitude for the expected absolute meteor magnitude, equation~\ref{eq:expm} provides yet another estimate of $M_\text{ref}$:
\begin{align}
    M_\text{ref} &\approx \frac{1}{\ln r} + \frac{1}{N} \sum_k M_k'
    \label{eq:mest2}
\end{align}
Here we use the index $k$ to emphasize that this sum is performed over all meteors rather than all cells.
This estimate depends only on the observed absolute magnitudes and the population index; it does not depend at all on the cell-specific limiting magnitudes. 

Equation~\ref{eq:expm} has two important uses. First, it can be compared with equation~\ref{eq:mn} to determine whether the limiting magnitudes have been computed correctly. Second, it may be useful for estimating the reference limiting magnitude associated with a published flux value when only meteor counts and magnitudes are provided.

\section{Inapplicability to visual rates}
\label{sec:mzhr}

The ECA method as presented in Section~\ref{sec:eca} assumes that every meteor that is bright enough to be detected in a given cell \emph{will} be detected. However, \cite{koschack90a} found that there is no hard upper limit on meteor magnitude when it comes to human observers; instead, the probability of detection varies with the brightness of the meteor and its angular offset from the observer's line of sight. As a result, Equations~\ref{eq:mref}, \ref{eq:mn}, and \ref{eq:mest2} cannot be applied to visual observations; we derive an alternative in this section.

Equations~\ref{eq:fm} and \ref{eq:zhr} can be combined:
\begin{align}
    f(M_\text{ref} \, | \, \text{ZHR}, \, r) &=
    \nonumber \\
    \frac{\text{ZHR}}{\text{2840~km}^2} &(r - 1.26) (r - 1.3)^{0.748} \, r^{M_\text{ref} - 6.5} \label{eq:zhrcombo}
\end{align}
If we then set the derivative of this equation with respect to $r$ to zero, we obtain:
\begin{align}
    M_\text{KR} &= 6.5 - \frac{r}{r-1.26} - 0.748 \, \frac{r}{r-1.3}
    \label{eq:mkr}
\end{align}
However, equation~\ref{eq:zhrcombo} does not account for observing conditions. If a visual observer is viewing a hazy or light-polluted sky, they may not be able to detect meteors down to magnitude +6.5. We seek a reference magnitude that takes this into account.

We have been unable to reproduce the derivation of equation~\ref{eq:zhr} from the information provided in \cite{koschack90a} and \cite{koschack90b}. We nevertheless have attempted to simulate a reasonable range of magnitudes that is consistent with their general approach. First, we constructed an empirical profile for airmass vs.\ elevation angle ($\phi$) by repeatedly evaluating the following integral for different choices of $\phi$:
\begin{align}
    X(\phi) &= \int_0^{\infty} \rho_\text{atm}(h(s, \phi)) \, d s \\
    h(s, \phi) &= \sqrt{R_\oplus^2 + s^2 + 2 R_\oplus s \sin \phi } - R_\oplus
\end{align}
where $R_\oplus$ is the Earth's radius, $\rho_\text{atm}$ is the air density, and the integration variable $s$ is distance from the observer. We performed this task using the ``StandardAtmosphereData'' function in Mathematica \citep{mathematica}, which defaults to the 1976 United States Standard Atmosphere model \citep{atm1976}. 

Next, we divided the visible sky into bins and calculated the elevation angle and distance corresponding to each bin, assuming that meteors ablate at a height of 100~km. The limiting magnitude of each bin is then:
\begin{align}
    M_i &= m_0 - 5 \ln \frac{d}{100~\text{km}} - X(\phi)
\end{align}
where $m_0$ is the visual limit, the dimmest magnitude of an object visible to the observer.

Although \cite{koschack90a} did not settle on a satisfactory functional form for the probability of perception, we were able to approximate Table~4 of \citeauthor{koschack90a} using the following formula:
\begin{align}
    p(\delta, \alpha) &= \text{expit} \lbrace - \left( a + b \, \delta + c \tan \frac{\alpha}{2} \right) \rbrace \, , ~ \text{where} \label{eq:perception} \\
    \delta &= m' - m_0 < 0
\end{align}
Here ``expit'' is the standard logistic function, $m'$ is apparent meteor magnitude, $\alpha$ is the angular separation between the observer's line of sight (LOS) and the meteor, and the constants are ${a = 2.6240}$, ${b = 1.3815}$, and ${c = 10.6136}$. This approximation gives us values that are typically within 0.05 of the correct fractional value. 

\begin{figure}
    \centering
    \includegraphics[width=0.9\linewidth]{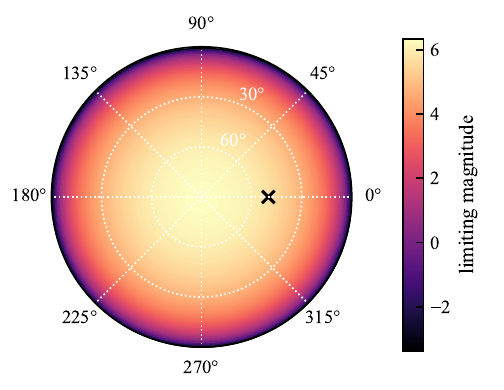}
    \includegraphics[width=0.9\linewidth]{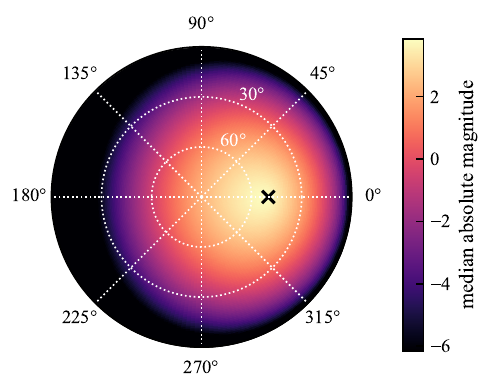}
    \includegraphics[width=0.9\linewidth]{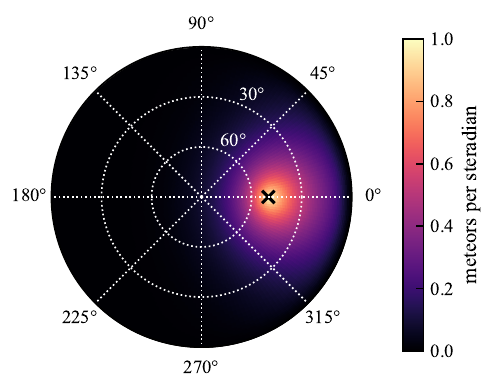}
    \caption{(Top) The limiting absolute meteor magnitude that can be seen as a function of the meteor's azimuth and elevation angle relative to a visual observer. The black ``x'' marks the direction of the observer's gaze, which has an elevation of 50$^\circ$. (Middle) The median absolute meteor magnitude that can be seen by the observer, assuming a population index of ${r=2.4}$. Even in the center of the observer's field of view, the median magnitude is several magnitudes brighter than the limiting magnitude; this is because observing a meteor near the limit, while possible, is very unlikely. (Bottom) The number of meteors observed per unit solid angle. This number has been scaled so that it is unity at the center of the observer's field of view; in reality it will depend on the shower's ZHR.}
    \label{fig:maps}
\end{figure}

We can see from Equation~\ref{eq:perception} that the probability of observing a meteor right at the limit of detectability (${\delta \sim 0}$) is quite low, even in the center of the observer's field of view (${\alpha = 0}$, ${p \simeq 0.07}$). The observer has a much better chance of observing meteors that are several magnitudes brighter than the detectability limit and that pass near the direction of their gaze. As a result, the typical magnitude observed will be quite a bit brighter than the limiting magnitude, especially in the peripheral visual field. This effect is illustrated in Figure~\ref{fig:maps}.

The probability of perception complicates the determination of the reference magnitude. The number of meteors we expect to see in a given area is now:
\begin{align}
    \langle n_i \rangle &\propto r^{M_i - M_\text{ref}} A_i \, P_i(r)
\end{align}
where $P_i$ is the fraction of all meteors we expect to see in that area:
\begin{align}
    P_i(r) &= \int_{-\infty}^0 p(\delta, \alpha_i) \cdot r^\delta \ln r \, d \delta 
    \label{eq:pint}
\end{align}
Because $P_i$ is a function of $r$, it must be included in our calculation of ${d A_\text{eff}/d r}$. Equation~\ref{eq:mref} is replaced with:
\begin{align}
    M_\text{ZHR} &= \frac{
        \sum_i M_i \, A_i \, r^{M_i} \, P_i - 
        r \sum_i A_i \, r^{M_i} \, \frac{d P_i}{d r}
    }{\sum_i A_i \, r^{M_i} \, P_i}
\end{align}
This equation must be evaluated numerically. We have done so, using a range of plausible $r$ values, and found that
\begin{align}
    M_\text{ZHR} - m_0 &\approx M_\text{KR} - 6.5
    \label{eq:mzhr0}
\end{align}
The two expressions agree to within 0.2 magnitudes between ${r=1.7}$ and ${r=3.0}$. Furthermore, we find that the average apparent magnitude can be approximated as follows:
\begin{align}
    \bar{m}' &\approx m_0 - 3.8 + 2.23 \ln (r-1.3) 
    \label{eq:barm}
\end{align}
This expression allows us to estimate the visual limit from the mean apparent magnitude. Assuming the classical +6.5 limit, the mean apparent magnitude will be approximately +2.9 for ${r = 2.4}$.

If we combine equations~\ref{eq:mkr}, \ref{eq:mzhr0}, and \ref{eq:barm}, we can estimate the reference magnitude directly from the mean apparent magnitude:
\begin{align}
    M_\text{ZHR} \approx \bar{m}' &+ 3.8 - 2.23 \ln (r-1.3) 
    \nonumber \\
    &- \frac{r}{r-1.26} - 0.748 \, \frac{r}{r-1.3}
    \label{eq:mzhr}
\end{align}
This approximate reference magnitude has both of our desired qualities: it minimizes the dependence of flux on $r$ and it is representative of the range of magnitudes observed. It does, however, require observers to record the apparent magnitudes of individual meteors; fortunately, the International Meteor Organization (IMO) Visual Meteor Database (VMDB) includes apparent magnitudes for a subset of all observations.

\begin{figure}
    \centering
    \includegraphics[width=\linewidth]{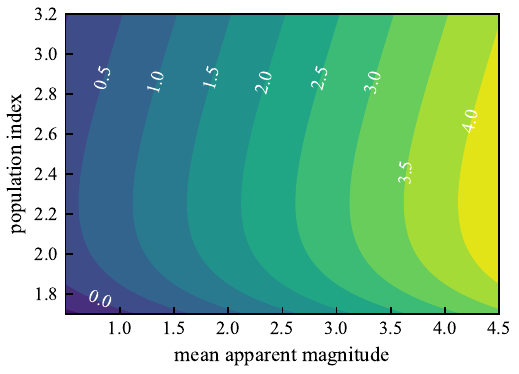}
    \caption{Approximate reference magnitude ($M_\text{ZHR}$) as a function of population index ($r$) and mean apparent magnitude ($\bar{m}'$), calculated using Equation~\ref{eq:mzhr}. For example, if ${\bar{m}' = 3}$ and ${r = 2.4}$, then $M_\text{ZHR}$ is between 2.5 and 3.}
    \label{fig:mzhr}
\end{figure}

We would like to point out that our treatment of ZHR and reference magnitude neglects the difference between visual and camera or photometric magnitude. For instance, \cite{jacchia57} found that visual magnitude estimates were 1-2 magnitudes brighter than photographic magnitude measurements for the same meteors. This difference, which they call the ``color index,'' is a function of photographic magnitude (see Figure~12b of that paper) and is attributed to the Purkinje effect \citep[see also][]{protte20}. If we assume that the perception probability applies to visual magnitude and use another sigmoid function to describe the color index, we predict an average apparent \emph{visual} magnitude that is about half a magnitude dimmer than Equation~\ref{eq:barm}. Unfortunately, Equation~\ref{eq:pint} becomes quite complicated in this case. A full treatment of this effect is beyond the scope of this paper; for now, we will simply consider the uncertainty associated with Equation~\ref{eq:mzhr} to be at least 0.5 magnitudes.

We began this section by drawing a distinction between the hard detection threshold assumed for instrumental detection and the lack of such a threshold for visual observations. However, instrumental detection thresholds will also be ``softened'' to some extent by noise. If the magnitude range of this softening effect is non-negligible compared to the range of detected magnitudes in the cell or the range of limiting magnitudes represented in the network, then the use of the ECA method -- and Equations~\ref{eq:mref}, \ref{eq:mn}, and \ref{eq:mest2} -- may not be appropriate.

\section{Application to the Orionids}
\label{sec:ori}

In this Section, we compare fluxes from four independent meteor observation networks and demonstrate the benefit of using our proposed reference magnitude. These networks are the Global Meteor Network \citep[GMN;][]{vida21}, the IMO Video Meteor Network \citep[VMN;][]{molau14}, the IMO Visual Meteor Database \citep[VMDB;][]{roggemans88}, and the Canadian Meteor Orbit Radar \citep[CMOR;][]{jones05}. 

We selected the Orionids for this demonstration because all four networks assume quite different values of $r$ for this shower (see Table~\ref{tab:networks}). Normally, this would prohibit a useful synthesis of the data. However, we will attempt to estimate the reference magnitude of each network and show that this removes some of the discrepancy in activity. We include the past four years of Orionid activity; this choice is determined by GMN, whose fluxes date back to 2020.

All three flux networks (GMN, VMN, and CMOR) use the same flux-to-ZHR conversion (eqs.~\ref{eq:fm} and \ref{eq:zhr}). Thus, once we determine $M_\text{ref}$ for each network, we can convert the published ZHR values for that network to $f(M_\text{ref})$ by inverting these equations, using each source's assumed $r$.

\begin{table}
    \caption{Orionid population index ($r$) and reference magnitude for four independent networks.}
    \begin{center}
    \begin{tabular}{lccc} \hline \hline
        network & $r$ & $M_\text{ref}$ & $M_\text{ZHR}$ \\ \hline
        CMOR & 3.0 & +8.0 \\
        GMN & 2.2 & +3.9 \\
        VMN & 2.5 & +2.0 \\
        VMDB & 2.4 & & +2.6
    \end{tabular}
    \label{tab:networks}
    \end{center}
    \textsc{Note} -- The population index ($r$) listed is the value used to calculate meteor flux or ZHR by each network. We also include our estimated reference magnitude ($M_\text{ref}$ or $M_\text{ZHR}$) for each network where applicable.
\end{table}

\subsection{Global Meteor Network (GMN)}

GMN meteor fluxes are already reported to the appropriate reference magnitude \citep[Equation~\ref{eq:mref}; these are called time-area-product or TAP-weighted magnitudes in][]{vida22}. In fact, the GMN flux data files report the reference magnitude for each observation interval as well as the overall average; the reference magnitude may vary over time due to changes in observing conditions \citep{vida22}. In Table~\ref{tab:networks}, we report the reference magnitude averaged over all years.

\newpage
\subsection{Video Meteor Network (VMN)}
\label{sec:vmn}

VMN fluxes can be obtained from a web portal\footnote{\url{https://meteorflux.org/}}; these fluxes are quoted to magnitude +6.5 and are also converted to ZHR. We used the default options for the Orionids, with one exception: we opted to turn on the perception coefficient (PC) correction.

\begin{figure}
    \centering
    \includegraphics{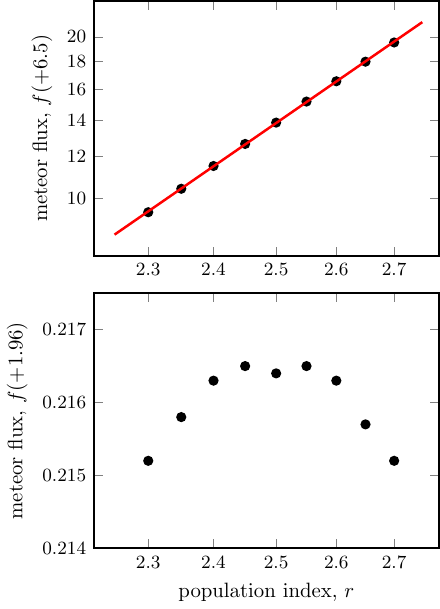}
    \caption{(Top) Magnitude-6.5-limited meteor flux output by the VMN web portal as a function of user-selected population index ($r$). Note that these data are presented on a log-log scale and the red line represents the best fit line between ${\ln f(+6.5)}$ and ${\ln r}$. (Bottom) Equivalent flux for  ${f(M_\text{ref}=+1.96)}$. Note that our calculated $M_\text{ref}$ minimizes the variation in $f(M_\text{ref})$ with respect to $r$.}
    \label{fig:vmn}
\end{figure}

The website does not report meteor magnitudes, but it does allow the user to vary the assumed population index \citep{molau14imc}. This gives us an opportunity to extract $M_\text{ref}$ using yet another approach. Using equation~\ref{eq:fm}, we calculate:
\begin{align}
    \frac{d \ln f(+6.5)}{d \ln r} &=
    \frac{r}{f(M_\text{ref})} \frac{d f(M_\text{ref})}{d r} + 6.5 - M_\text{ref}
\end{align}
Because ${d f(M_\text{ref})/d r = 0}$, we can therefore obtain $M_\text{ref}$ by repeatedly changing the assumed value of $r$ and regressing the reported ${\ln f(+6.5)}$ against ${\ln r}$. The results of this process are shown in the top panel of Figure~\ref{fig:vmn}. We obtain ${M_\text{ref} = 1.96}$ in this manner. The bottom panel of Figure~\ref{fig:vmn} demonstrates that this value does in fact minimize the variation in flux; when we use $M_\text{ref}$, the variation in flux over ${r = 2.3}$ to ${r = 2.6}$ is less than 1\%, which is much less than the factor-of-two range we see in $f(+6.5)$. Figure~\ref{fig:vmn} also indicates that there is some noise in the data, which is probably due to rounding error; we therefore round our reference magnitude to +2.

\subsection{Visual Meteor Database (VMDB)}

Zenithal hourly rates (ZHR) for major showers are published by the International Meteor Organization (IMO) on their website.\footnote{\url{https://www.imo.net/members/imo_live_shower}} These ZHRs are automatically computed from user-supplied data and the IMO does not recommend their use for scientific applications. We ignored this warning for the purposes of this demonstration and downloaded Orionid ZHR values for the years 2020 through 2023. The IMO uses ${r = 2.4}$ for the ``whole campaign.'' 

Magnitude counts are available to IMO members via the Video Meteor Database (VMDB).\footnote{\url{https://www.imo.net/members/imo_vmdb}} We downloaded these data for Orionids between 2020 and 2023 (inclusive). Because visual observations do not contain distance information, these are apparent magnitudes; we found that the mean apparent magnitude was +2.75. When we input this value, along with ${r=2.4}$ into equation~\ref{eq:mzhr}, we obtain a reference magnitude of ${M_\text{ZHR} = 2.6}$.

\subsection{Canadian Meteor Orbit Radar (CMOR)}

CMOR is a patrol radar; it detects the reflection of an emitted radar pulse off of the surface of the cylinder of plasma produced by an ablating meteor. The ability to detect a meteor is therefore not determined by its brightness, but rather by the electron line density of the trail (usually denoted $q$). Furthermore, CMOR does not necessarily detect the meteor when its trail is at its peak electron line density. Instead, it detects a meteor when its velocity has no radial component relative to the radar station and its trail is therefore perpendicular to the radar beam.

Electron line density can be converted to an equivalent radar magnitude, but individual radar magnitudes are not readily available for those meteors contributing to the Orionid fluxes measured by CMOR. However, the CMOR flux pipeline accepts population index (in the form of a mass index) as an input parameter.  We therefore varied $r$ in order to extract $M_\text{ref}$ using the approach outlined in \ref{sec:vmn}. We found that ${M_\text{ref} = +8.0}$, which is fairly close to the limiting magnitude of +8.4 quoted in \cite{cb04}.

\subsection{Synthesis}

\begin{figure}
    \centering
    \includegraphics{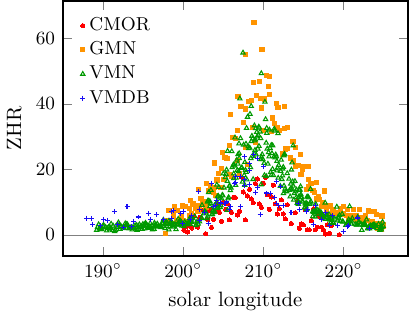}
    \caption{Raw Orionid ZHR values from four different networks recorded between 2020 and 2024.}
    \label{fig:rawzhr}
\end{figure}

The raw ZHR values reported for each network are shown in Figure~\ref{fig:rawzhr}. The peak values are quite discrepant; the CMOR and VMDB data have a peak ZHR of approximately 15, the VMN data have a peak ZHR of about 30, and the GMN data have a peak ZHR of about 50.

\begin{figure}
    \centering
    \includegraphics{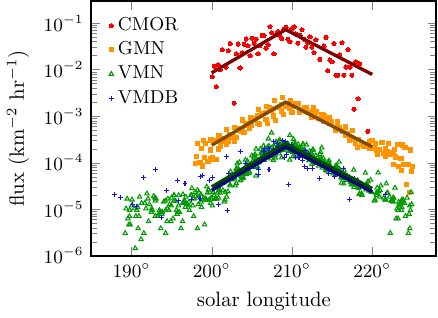}
    \caption{Background-subtracted Orionid fluxes from four networks; these fluxes correspond to the reference magnitudes listed in Figure~\ref{tab:networks}. The dark line superimposed on each data set represents our fitted profile.}
    \label{fig:fref}
\end{figure}

We convert ZHR to flux using equation~\ref{eq:zhrcombo} and each network's chosen value of $r$. We also fit a double-exponential function \citep{jenniskens94} to the data in order to extract the peak value. Some of the data in Figure~\ref{fig:rawzhr} show signs of sporadic contamination; for instance, both video networks appear to asymptote around ${\text{ZHR} = 1}$. We therefore simultaneously fit for a constant background level and subtract it from the flux. The results are shown in Figure~\ref{fig:fref}.

\begin{figure}
    \centering
    \includegraphics{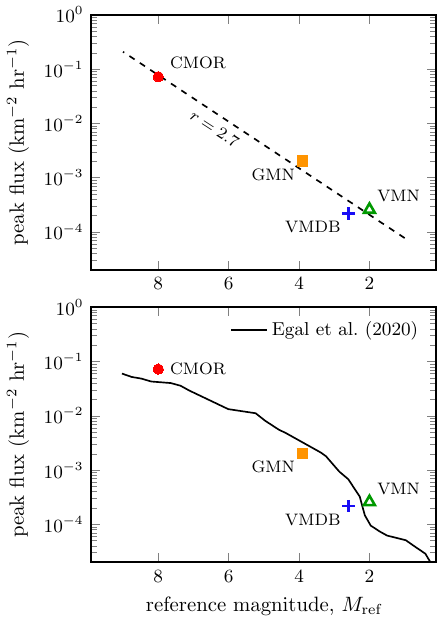}
    \caption{The peak flux of the Orionid meteor shower from four networks; each point on this plot corresponds to the maximum value of a fitted flux profile in Figure~\ref{fig:fref}. In the top panel, we compare the data with a constant population index of ${r=2.7}$; in the bottom panel, we compare the same data with the mass-magnitude profile of \cite{egal20}.}
    \label{fig:fvsm}
\end{figure}

Our profile fits allow us to extract the peak ZHR or flux from each data set. These peak fluxes are then plotted against the reference magnitudes in Figure~\ref{fig:fvsm}. If the mass distribution does indeed take the shape of a power law, then this plot should resemble a line. If we fit for $r$ using these four points, we obtain an overall population index of ${r = 2.7}$ (corresponding to a peak ZHR of 28). On the other hand, the Orionid mass distribution may not resemble a power law over large ranges in mass. The dynamical simulations of \cite{egal20} predict a relative lack of Orionid particles at sizes that are roughly comparable with CMOR's limiting magnitude. In the bottom panel of Figure~\ref{fig:fvsm}, we compare our peak fluxes with the relative particle counts shown in Figure~13 of \cite{egal20}, extracted using WebPlotDigitizer \citep{rohatgi22}. Masses have been converted into magnitudes using the mass-magnitude relation of \cite{verniani73}, and an arbitrary scaling has been applied to the particle counts to bring them into rough agreement with the data. Both distributions appear to be plausible matches to the data.

\begin{figure}
    \centering
    \includegraphics{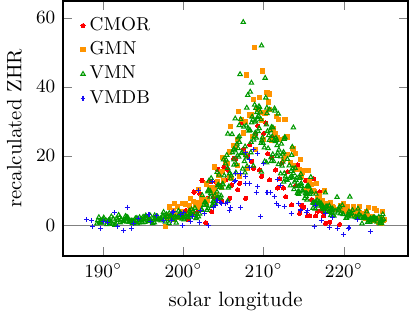}
    \caption{Orionid ZHR values from all four networks, calculated from $f(M_\text{ref})$ or $f(M_\text{ZHR})$ using a uniform population index of ${r = 2.7}$. We have subtracted the estimated background contamination. }
    \label{fig:corr}
\end{figure}

Although we have obtained ${r = 2.7}$ using the measurements discussed in this paper, we do not suggest that the meteor astronomy community should necessarily adopt this value for the Orionid meteor shower. Our intent is instead to demonstrate how our proposed reference magnitude can be used to better compare disparate results. When fluxes are quoted to a network-specific reference magnitude, we can combine independent measurements to probe the magnitude distribution and reveal any remaining discrepancies. For instance, if we convert flux back to ZHR using ${r = 2.7}$ for all networks, we find that the discrepancy between networks has been reduced (see Figure~\ref{fig:corr}), but not eliminated. The spread in peak ZHR values has been reduced by approximately one-third (see Table~\ref{tab:corr}). The visual rates are still lower than the instrumentally measured rates but, because we have accounted for differences in the assumed value of $r$, we now know that this discrepancy must be due to some other factor.

\begin{table}
    \caption{Peak ZHR values from each network (second column) and after correcting for differences in $r$ (third column).}
    \label{tab:corr}
    \begin{center}
    \begin{tabular}{l|cc} \hline \hline
        network & ZHR & ZHR$_\text{corr}$ \\ \hline
        CMOR & 15.3 & 25.7 \\ 
        GMN  & 51.3 & 41.1 \\
        VMN  & 31.8 & 34.3 \\
        VMDB & 15.3 & 16.0
    \end{tabular}
    \end{center}
\end{table}

\newpage
\section{Conclusion}
Currently, meteor fluxes are typically scaled to ZHR or a standard limiting magnitude of +6.5 prior to publication. We argue that fluxes should be reported to a network-specific reference magnitude that is equal to a weighted average of the limiting magnitude of the observing station(s). This choice of reference magnitude minimizes the dependence of the reported flux on the assumed population index and therefore enables the synthesis of independent meteor shower flux measurements. This is particularly useful when the population index varies with magnitude, as the index that is most appropriate for computing flux from a particular set of observations may differ from the average index over a large range of magnitudes.

If researchers choose to publish fluxes using our suggested reference magnitude, we suggest that they also include some measure of the degree of variation within their network's limiting magnitude. We have suggested the sample standard deviation of the limiting magnitude. If the variation in limiting magnitude is very large and if there are a sufficient number of meteor detections, the researcher could also consider subdividing the data into magnitude ranges \citep[e.g.,][]{molau14imc}.

In many cases, fluxes are quoted to a minimum mass rather than limiting magnitude. We favor the use of a reference magnitude rather than a reference mass for two reasons. First, magnitude is more closely related to the observed quantity: the apparent brightness seen by the observer or observing station. This is reflected in the effective collection area method for measuring fluxes, which depends only on limiting magnitude. The second reason is that there are multiple competing formulae for converting magnitude to mass \citep{jvb67,verniani73,cb16}, and therefore the conversion of magnitude to mass creates another opportunity to introduce discrepancies that arise from different assumptions rather than real differences in the data.

To demonstrate the utility of our reference magnitude, we estimated its value for several independent measurements of Orionid fluxes between 2020 and 2024. We included data from four networks, each of which used a different assumed population index to compute meteor flux or ZHR. We found that the data could be brought into closer agreement if we assume that the Orionid mass index is approximately 2.7. This synthesis could potentially be improved if exact, rather than estimated, reference magnitudes for CMOR and VMN were to be published. The data also suggest that a meteor observation network of extremely sensitive cameras with a reference magnitude of approximately +6 \citep[similar to electron-multiplied charge-coupled device, or EMCCD, cameras;][]{brown2020coordinated, mills21, Gural2022} would be a valuable addition.
\vspace{\baselineskip}

\noindent
This work was supported in part by NASA Cooperative Agreement 80NSSC21M0073 and by the Natural Sciences and Engineering Research Council of Canada.

\bibliography{refs}{}
\bibliographystyle{aasjournal}

\end{document}